\documentclass[prd,twocolumn,aps,superscriptaddress,nofootinbib,tightenlines]{revtex4}

\usepackage{amsfonts}
\usepackage{braket}
\usepackage{bm}
\usepackage{pstricks,pst-node}

{\count255=\time\divide\count255 by 60 \xdef\hourmin{\number\count255}
  \multiply\count255 by-60\advance\count255 by\time
  \xdef\hourmin{\hourmin:\ifnum\count255<10 0\fi\the\count255}}

\def\nn{ \nonumber \\ }
\def\abs#1{\left| #1 \right|}
\def\ng{\mathcal{N}}
\def\loop#1{\Braket{ #1 }}
\def\chain#1{\{  #1 \} }
\def\uchain#1{(  #1 ) }

\begin{document}

\title{Rephasing Invariants of Quark and Lepton Mixing Matrices}

\author{Elizabeth~Jenkins}

\affiliation{University of California, San Diego, 9500 Gilman Drive, La Jolla, CA 92093}

\author{Aneesh V.~Manohar}

\affiliation{University of California, San Diego, 9500 Gilman Drive, La Jolla, CA 92093}

\date{\today}

\begin{abstract}
Rephasing invariants of quark and lepton mixing matrices are obtained in the standard model extended by the seesaw mechanism, and in its low-energy effective theory with the dimension-five Majorana mass operator. We classify the basic invariants, discuss non-trivial relations between them, and determine the independent invariants which characterize all the information in the mixing matrices in a basis-independent way. We also discuss the restrictions on the allowed ranges for the mixing phases, and on the rephasing invariants, which follow from a discrete invariance of the Majorana mass matrix.
\end{abstract}

\maketitle

\section{Introduction}

The interaction of quarks with weak gauge bosons is described in terms of the Cabibbo-Kobayashi-Maskawa (CKM) mixing matrix $V$, a unitary $3 \times 3$ matrix, which is the transformation matrix between the quark mass eigenstate and weak interaction eigenstate bases. In the quark mass eigenstate basis, one still has the freedom to make phase rotations on the quark fields, which leads to the redefinition
\begin{eqnarray}
V &\to& e^{i \Phi_U} V e^{-i \Phi_D} 
\label{ckm}
\end{eqnarray}
where $\Phi_U=\text{diag}(\phi_u,\phi_c,\phi_t)$ and  $\Phi_D=\text{diag}(\phi_d,\phi_s,\phi_b)$. Physical quantities are basis independent, and must be invariant under the rephasing Eq.~(\ref{ckm}). CKM rephasing invariants have been studied extensively in the literature~\cite{jarlskog,greenberg,dunietz}, the best-known example being the $CP$-odd Jarlskog invariant $J=\text{Im}\, V_{11}V_{22}V_{12}^* V_{21}^*$. 

In this paper, we extend the analysis of rephasing invariants to the lepton sector. We will assume that neutrino masses in the lepton sector are described by the seesaw mechanism. At low energies (below the seesaw scale), the theory reduces to the standard model with an additional dimension-5 operator which leads to Majorana masses for the weak-doublet neutrino fields after electroweak symmetry breakdown. The mixing matrix in this case is the PMNS matrix $U$. Rephasing invariants of $V$ and $U$ have previously been studied by Nieves and Pal~\cite{np1,np2}. We review their analysis, and give additional results on relations between rephasing invariants. We also characterize the invariants in a different way, which gives a better understanding of the independent invariants, and how they encode the information contained in the mixing matrices.
At energies above the seesaw scale, the lepton sector has two mixing matrices $V$ and $W$; we extend the analysis of rephasing invariants to this case. We will see that the classification of lepton rephasing invariants is considerably more involved than the quark invariants, and that there are many non-trivial relations between the invariants. These relations provide additional insight into the structure of $CP$-violating observables in the extended standard model.

The high-energy lepton invariants are not all currently accessible experimentally.  Some of them (particularly the $CP$-odd ones) are relevant for leptogenesis, and so can be constrained indirectly~\cite{leptogenesis1,leptogenesis2}.  Many of the low-energy lepton invariants are already measured in neutrino oscillation experiments.  The remaining unmeasured low-energy lepton invariants are the subject of the current and future program of neutrino oscillation and neutrinoless double beta decay experiments.
The relation between the high-energy and low-energy invariants is particularly interesting, and will be discussed in a future publication~\cite{jm}.

There is an extensive literature on quark and lepton invariants~(see, e.g. \cite{kusenko,susy,Branco1,Branco2}). In addition to rephasing invariants discussed here, one can construct invariants directly from products of the the quark and lepton mass matrices, rather than from the mixing matrices. These invariants are often called weak basis invariants~\cite{Branco1,Branco2} and will be discussed in Ref.~\cite{jm}.

The paper is organized as follows. In Sec.~\ref{sec:mixing}, we define the high-energy and low-energy theories, their mixing matrices, and rephasing transformations of the mixing matrices. In Sec.~\ref{sec:parameter}, we write the mixing matrices in a standard form, review the counting of independent parameters and the allowed ranges for the angles and phases. Section~\ref{sec:rephasing} classifies the quark invariants and the lepton invariants in the low-energy and high-energy theories, and discusses their properties. The conclusions are given in Sec.~\ref{sec:conclusions}.

\section{Mixing Matrices}\label{sec:mixing}

Neutrino masses can be included in the standard model by introducing singlet left-handed neutrino fields $N^c$.\footnote{The left-handed singlet neutrino fields are denoted by $N^c$ because the usual convention in the literature is to denote the right-handed field by $N$.}  The relevant terms in the seesaw Lagrangian are the Yukawa couplings and singlet Majorana mass terms,\begin{eqnarray}
{\cal L} &=& -  E^c_i \left(Y_E\right)_{ij}  L_j H^\dagger 
 -   N^c_i \left(Y_\nu\right)_{ij}  L_j H\nn
 &&-\frac 1 2 N^c_i M_{ij} N^c_j +  \text{h.c.} ,
\label{high}
\end{eqnarray}
where $H$ is the Higgs field, all fermions fields are left-handed, $i,j$ are flavor indices, and gauge and Lorentz indices have been suppressed. The seesaw Lagrangian violates lepton number if \emph{both} $Y_\nu \not=0$ and $M \not =0$. We study the general case of $\ng$ generations, and later restrict to the physical case of interest $\ng=3$. In this paper, we assume, for simplicity, that the number of singlet neutrinos is equal to the number of lepton doublets. The generalization to a different number is straightforward~\cite{bgj1}. Flavor mixing in the lepton sector is governed by the Yukawa matrices $Y_{E}$ and $Y_\nu$, which are $\ng \times \ng$ complex matrices, and by the mass matrix $M$, which is a $\ng \times \ng$ complex symmetric matrix. Under $CP$, $Y_{E} \to Y^*_{E}$, $Y_\nu \to Y^*_\nu$ and $M \to M^*$.

Integrating out the $N^c$ fields leads to the effective Lagrangian below the seesaw scale~\cite{bgj1,bgj2},
\begin{eqnarray}
{\cal L} &=& -E^c_{i} \left(Y_E\right)_{ij} L_j  H^\dagger 
+ \frac12 \left( L_{i} H \right) \left(C_5\right)_{ij} \left( L_{j} H \right)+\text{h.c},\nn
\label{low}
\end{eqnarray}
where at lowest order, the dimension-5 coefficient~\cite{weinberg}
\begin{eqnarray}
C_5 = Y_\nu^T M^{-1} Y_\nu
\end{eqnarray}
is an $\ng \times \ng$ complex symmetric matrix. Under $CP$, $C_5 \to C_5^*$. The neutrino Majorana mass matrix in the low energy theory is $-C_5 v^2/2$, where $v\sim 247$~GeV is the Higgs vacuum expectation value.

Unitary field redefinitions on $L$, $E^c$ and $N^c$ can be used to diagonalize the matrices $Y_E$ and $M$ in the seesaw Lagrangian Eq.~(\ref{high}), 
\begin{eqnarray}
Y_E \to \Lambda_E,\qquad
M \to \Lambda_N,
\end{eqnarray} 
where $\Lambda_{E,N}$ are diagonal matrices with real, non-negative eigenvalues. In this basis, the matrix $Y_\nu$ can be written as
\begin{eqnarray}
Y_\nu &=& W^{-1} \Lambda_\nu V,
\end{eqnarray}
where $V$ and $W$ are unitary matrices, and $\Lambda_\nu$ is a diagonal matrix with real, non-negative eigenvalues. $V$ and $W$ are the mixing matrices which describe flavor violation in the lepton sector of the seesaw theory.   $V$ is the analogue of the CKM quark mixing matrix in the lepton sector;\footnote{The same symbol $V$ is used for both matrices. It should be clear from the context whether we are referring to the quark or lepton mixing matrix.}\ it describes the mismatch between the unitary field redefinitions on $E$ and $\nu$ in $L$ required to diagonalize $Y_E$ and $Y_\nu$, respectively.  $W$ is an additional mixing matrix which has no quark analogue; it describes the mismatch between the unitary field redefinitions on $N^c$ required to diagonalize $M$ and $Y_\nu$, respectively.  

Arbitrary rephasing transformations on the weak doublet neutrino fields $\nu$ are allowed because the phase redefinitions  $V \to  e^{i \Phi_\nu} V$, $W \to  e^{i \Phi_\nu} W$, where $\Phi_\nu$ is diagonal and real, leave $Y_\nu$ invariant since $e^{-i \Phi_\nu} \Lambda_\nu e^{i \Phi_\nu} = \Lambda_\nu$.  There also is the freedom to: (a) make the \emph{same} diagonal rephasing transformations on $L$  and $E^c$, which leaves $\Lambda_E$ invariant, and (b) multiply $N^c$ fields by $-1$, which leaves 
$\Lambda_N$ invariant. The full rephasing transformation is
\begin{eqnarray}
V \to e^{i \Phi_\nu} V e^{-i \Phi_E},\qquad
W \to e^{i \Phi_\nu} W \eta_N,
\label{rph}
\end{eqnarray}
where $\eta_N$ is a diagonal matrix with allowed eigenvalues $\pm1$.  The $\eta_N$ matrix takes into account $-1$ rephasings allowed for the $N^c$ fields.

In the effective theory, unitary transformations on $E^c$ and $L$ can be used to diagonalize $Y_E$ in Eq.~(\ref{low}), $Y_E \to \Lambda_E$,  where $\Lambda_E$ is a diagonal matrix with real, positive eigenvalues.  In this basis, $C_5$ can be written as
\begin{eqnarray}
C_5 &=& U \Lambda_5 U^T \ ,
\label{c5diag}
\end{eqnarray}
where $\Lambda_5$ is a diagonal matrix with real, non-negative eigenvalues.  Eq.~(\ref{c5diag}) defines the unitary PMNS matrix $U$ which diagonalizes the effective Majorana mass matrix of the weakly interacting neutrinos.  This mixing matrix is responsible for neutrino oscillations in low-energy experiments. The rephasing transformation for the PMNS matrix is
\begin{eqnarray}
U &\to& e^{-i \Phi_E} U \eta_\nu,
\label{rpl}
\end{eqnarray}
where $\eta_\nu$ is a diagonal matrix with allowed eigenvalues $\pm1$.  The $\eta_\nu$ matrix takes into account $-1$ rephasings allowed for the Majorana $\nu$ fields.  

In the following sections, we will discuss invariants in the high-energy theory with Lagrangian Eq.~(\ref{high}) built out of $V,W$ and in the low-energy theory with Lagrangian Eq.~(\ref{low}) built out of $U$.  One does not need to consider invariants built out of all three matrices, since $U$ and $\{W,V\}$ do not exist in the same theory.

\section{Parameter Counting}\label{sec:parameter}

An $\ng \times \ng$ unitary matrix has $\ng^2$ parameters which are divided into $CP$-even and $CP$-odd parameters called angles and phases, respectively; there are $\ng(\ng-1)/2$ angles and $\ng(\ng+1)/2$ phases. As is well-known, the rephasing invariance Eq.~(\ref{ckm}) removes $(2\ng-1)$ phases from the quark mixing matrix $V$.  (An overall phase transformation with $\Phi_U = \Phi_D \propto \openone$ leaves $V$ invariant, and does not correspond to a removeable phase.)  Thus, $V$ can be rewritten in the standard CKM form with $\ng(\ng-1)/2$ angles and $(\ng-1)(\ng-2)/2$ phases. The angles $\theta_i \in [0, \pi/2]$, and the phases $\delta_i \in [0,2\pi)$.

Many different parameterizations of the CKM matrix have been discussed in the literature~\cite{ckmparam}. The particular form chosen is not important. We will pick a standard form $\mathcal{V}(\theta_i,\delta_i)$ which is a \emph{fixed} functional form. The quark CKM matrix $V$ and the lepton mixing matrices $U$, $V$ and $W$ will be given in terms of $\mathcal{V}$ by choosing (different) values for $\theta_i,\delta_i$ for each matrix. An arbitrary unitary matrix can then be written as
\begin{eqnarray}
e^{i\chi} e^{i \Phi} \mathcal{V}(\theta_i,\delta_i) e^{i \Psi}
\label{std}
\end{eqnarray}
where $\chi$ is an overall phase, $\Phi=\text{diag}(0,\phi_2,\ldots, \phi_{\ng})$,  and $\Psi=\text{diag}(0,\psi_2,\ldots, \psi_{\ng})$. The phases $\chi$, $\phi_i$, and $\psi_i$, $i=2,\ldots,\ng$, are the $(2 \ng -1)$ phases which can be removed in the CKM matrix by the rephasing transformation Eq.~(\ref{ckm}).   The
standard parameterization of the CKM matrix is
$V=\mathcal{V}(\theta_i,\delta_i)$.

In the low-energy effective theory, there is only one mixing matrix, the PMNS matrix $U$, with the rephasing invariance Eq.~(\ref{rpl}). Starting with the canonical form for the unitary matrix Eq.~(\ref{std}), the $\ng$ arbitrary phases $\chi$ and $\phi_i$, $i = 2, \ldots, \ng$ can be eliminated using the $\ng$ phase redefinitions $\Phi_E$ in Eq.~(\ref{rpl}). The $\eta_\nu$ factor of Eq.~(\ref{rpl}) can be removed by $e^{i \psi_i} \to - e^{i \psi_i}$ if $(\eta_\nu)_{ii}=-1$. [If $(\eta_\nu)_{11}=-1$, the redefinition $\eta_\nu \to -\eta_\nu$, $\chi \to \chi+\pi$  transfers the sign to the removeable phase $\chi$.] Thus the $\eta_\nu$ rephasing in Eq.~(\ref{rpl}) implies that the phases $\psi_i$ and 
$\psi_i+\pi$ are equivalent, so the range of the $\psi_i$ can be restricted to $\psi_i \in [0, \pi)$. It is convenient to have all phases vary over the same range $[0, 2\pi)$, so the standard form for the PMNS matrix rescales the $\psi_i$ phases by a factor of $1/2$ and is given by 
\begin{eqnarray}
U&=&\mathcal{V}(\theta_i,\delta_i) e^{i \Psi^/2}
\label{Ustd}
\end{eqnarray}
with $\ng(\ng-1)/2$ angles $\theta_i \in [0,\pi/2]$ and $\ng(\ng-1)/2$ phases consisting of $(\ng-1)(\ng-2)/2$ phases $\delta_i$ and $(\ng -1)$ phases $\psi_i$,  with range $\delta_i,\psi_i \in [0,2\pi)$.  For $\ng =3$, the low-energy mixing matrix $U$ has 3 angles and 3 phases.  We will call these parameters $\theta^{(U)}_{1,2,3}$, $\delta^{(U)}$, and $\psi^{(U)}_{2,3}$.

In the high-energy seesaw theory, there are two lepton mixing matrices $V$ and $W$, which can be written in the canonical form Eq.~(\ref{std}),
\begin{eqnarray}
V = e^{i\chi} e^{i \Phi} \mathcal{V}(\theta_i,\delta_i) e^{i \Psi},\quad
W = e^{i\chi^\prime} e^{i \Phi^\prime} \mathcal{V}(\theta_i^\prime,\delta_i^\prime) e^{i \Psi^\prime}.
\label{vwzero}
\end{eqnarray}
The rephasing transformations $\Phi_\nu$, $\Phi_E$ and $\eta_N$ of Eq.~(\ref{rph}) can be used to (i) eliminate $\chi$, $\chi^\prime$ and $\psi_i$, (ii) restrict $\psi^\prime_i$ to the range $[0,\pi)$ rather than $[0,2\pi)$, and (iii) eliminate \emph{either} $\Phi$ \emph{or} $\Phi^\prime$, \emph{but not both}.  For example, the standard form of the mixing matrices which uses the $\Phi_\nu$ phases to eliminate the $\Phi$ phases from $V$ is given by\footnote{Once again, it is convenient to rescale $\Psi^\prime \to \Psi^\prime/2$ so that all phases have the range $[0,2\pi)$.}
\begin{eqnarray}
V = \mathcal{V}(\theta_i,\delta_i), \qquad
W =  e^{-i \bar \Phi} \mathcal{V}(\theta_i^\prime,\delta_i^\prime) e^{i \Psi^\prime/2},
\label{vwone}
\end{eqnarray}
whereas the standard form of the mixing matrices which uses the $\Phi_\nu$ phases to eliminate the $\Phi^\prime$ phases from $W$ is given by
\begin{eqnarray}
V = e^{i \bar \Phi} \mathcal{V}(\theta_i,\delta_i), \qquad
W =  \mathcal{V}(\theta_i^\prime,\delta_i^\prime) e^{i \Psi^\prime/2}.
\label{vwtwo}
\end{eqnarray}
In Eq.~(\ref{vwone}), 
$V$ has the canonical CKM form with $\ng(\ng-1)/2$ angles $\theta_i$ and $(\ng-1)(\ng-2)/2$ phases $\delta_i$, whereas in Eq.~(\ref{vwtwo}), $W$ has the canonical PMNS form with $\ng(\ng-1)/2$ angles $\theta^\prime_i$ and  $\ng(\ng-1)/2$ phases consisting of the $(\ng-1)(\ng-2)/2$ phases $\delta^\prime_i$ and the $(\ng-1)$ phases $\psi^\prime_i$.  In either basis, there are $(\ng-1)$ additional phases $\bar \Phi \equiv \Phi -\Phi^\prime$ which cannot be removed, so there are a total of $\ng(\ng-1)$ phases between the two matrices.  Together, the matrices $V$ and $W$ contain $\ng (\ng-1)$ angles and $\ng (\ng-1)$ phases, for a total of $2\ng(\ng-1)$ parameters. 
Note that interactions involving only the $\nu$ and charged lepton fields can be written in terms of $V$ alone, whereas interactions involving only $\nu$ and $N^c$ fields can be written in terms of $W$ alone, so these processes do not depend on the $(\ng-1)$ phases in $\bar \Phi$.  The $\bar \Phi$ phases only enter in processes which depend on both $V$ \emph{and} $W$ involving $\nu$, $N^c$ and charged lepton fields.
Leptogenesis depends on the matrices $Y_\nu Y_\nu^\dagger$ and $M$~\cite{leptogenesis1,leptogenesis2}, and so is dependent only on $W$ and the $M$ mass eigenvalues, and is independent of $V$ and $\bar\Phi$.

For $\ng=3$, $V$ in canonical CKM form has 3 angles and 1 phase, which we will call $\theta_{1,2,3}^{(V)}$ and $\delta^{(V)}$, while $W$ in canonical PMNS form has 3 angles and 3 phases, which we will call $\theta_{1,2,3}^{(W)}$, $\delta^{(W)}$ and $\psi_{2,3}^{(W)}$.  When both matrices $V$ and $W$ are considered together, there are 2 additional phases which can be included in \emph{either} $V$ \emph{or} $W$.  We will call these phases $\bar \phi_{2,3}$.

\section{Rephasing Invariants}\label{sec:rephasing}

We now determine the rephasing invariants made out of the quark and lepton mixing matrices. There are three cases to consider: (a) quark invariants made out of the CKM matrix $V$ with the rephasing invariance Eq.~(\ref{ckm}); (b) lepton invariants in the low-energy theory made out of the PMNS matrix $U$ with rephasing invariance Eq.~(\ref{rpl}); and (c) lepton invariants in the high-energy theory made out of the lepton mixing matrices $V$ and $W$ with rephasing invariance Eq.~(\ref{rph}). 

It is easy to construct the invariants using a graphical analysis. The lepton mixing matrices are shown in Fig.~\ref{fig:1} for the high-energy theory, and in Fig.~\ref{fig:2} for 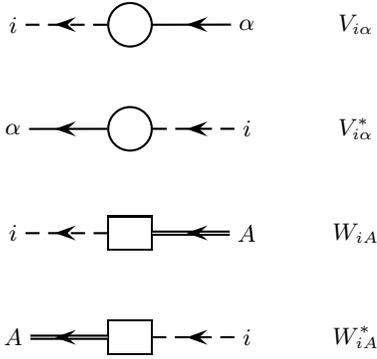
\begin{figure}
\begin{center}
\begin{pspicture}(0,0.5)(4.5,1.5)
\rput(4.5,1){$V_{i\alpha}$}
\rput(1.5,1){\circlenode{V}{$\phantom{V}$}}
\rput(3,1){\rnode{R}{$\ \alpha$}}
\rput(0,1){\rnode{L}{$i\ $}}
\ncline{R}{V}
\ncline[linestyle=dashed]{V}{L}
\psline[arrowsize=0.2]{->}(2.3,1)(2.25,1)
\psline[arrowsize=0.2]{->}(0.55,1)(0.5,1)
\end{pspicture}
\end{center}
\begin{center}
\begin{pspicture}(0,0.5)(4.5,1.5)
\rput(4.5,1){$V_{i\alpha}^*$}
\rput(1.5,1){\circlenode{V}{$\phantom{V}$}}
\rput(3,1){\rnode{R}{$\ i$}}
\rput(0,1){\rnode{L}{$\alpha\ $}}
\ncline[linestyle=dashed]{R}{V}
\ncline{V}{L}
\psline[arrowsize=0.2]{->}(2.3,1)(2.25,1)
\psline[arrowsize=0.2]{->}(0.55,1)(0.5,1)
\end{pspicture}
\end{center}
\begin{center}
\begin{pspicture}(0,0.5)(4.5,1.5)
\rput(4.5,1){$W_{i A}$}
\rput(1.5,1){\rnode{V}{\framebox{$\phantom{W}$}}}
\rput(3,1){\rnode{R}{$\ A$}}
\rput(0,1){\rnode{L}{$i\ $}}
\ncline[doubleline=true,doublesep=0.5\pslinewidth]{R}{V}
\ncline[linestyle=dashed]{V}{L}
\psline[arrowsize=0.2]{->}(2.3,1)(2.25,1)
\psline[arrowsize=0.2]{->}(0.55,1)(0.5,1)
\end{pspicture}
\end{center}
\begin{center}
\begin{pspicture}(0,0.5)(4.5,1.5)
\rput(4.5,1){$W_{i A}^*$}
\rput(1.5,1){\rnode{V}{\framebox{$\phantom{W}$}}}
\rput(3,1){\rnode{R}{$\ i$}}
\rput(0,1){\rnode{L}{$A\ $}}
\ncline[linestyle=dashed]{R}{V}
\ncline[doubleline=true,doublesep=0.5\pslinewidth]{V}{L}
\psline[arrowsize=0.2]{->}(2.3,1)(2.25,1)
\psline[arrowsize=0.2]{->}(0.55,1)(0.5,1)
\end{pspicture}
\end{center}
\caption{\label{fig:1} Graphical representation of elements of the mixing matrices in the high-energy theory. The dashed, solid and double lines are $\nu$, $E$ and $N^c$ indices, respectively.}
\end{figure}
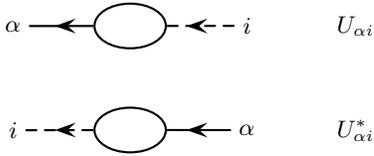
\begin{figure}
\begin{center}
\begin{pspicture}(0,0.5)(4.5,1.5)
\rput(4.5,1){$U_{\alpha i}$}
\rput(1.5,1){\ovalnode{V}{$\phantom{U_{\alpha i}}$}}
\rput(3,1){\rnode{R}{$\ i$}}
\rput(0,1){\rnode{L}{$\alpha\ $}}
\ncline[linestyle=dashed]{R}{V}
\ncline{V}{L}
\psline[arrowsize=0.2]{->}(2.3,1)(2.25,1)
\psline[arrowsize=0.2]{->}(0.55,1)(0.5,1)
\end{pspicture}
\end{center}
\begin{center}
\begin{pspicture}(0,0.5)(4.5,1.5)
\rput(4.5,1){$U_{\alpha i}^*$}
\rput(1.5,1){\ovalnode{V}{$\phantom{U_{\alpha i}}$}}
\rput(3,1){\rnode{R}{$\ \alpha$}}
\rput(0,1){\rnode{L}{$i\ $}}
\ncline{R}{V}
\ncline[linestyle=dashed]{V}{L}
\psline[arrowsize=0.2]{->}(2.3,1)(2.25,1)
\psline[arrowsize=0.2]{->}(0.55,1)(0.5,1)
\end{pspicture}
\end{center}
\caption{\label{fig:2} Graphical representation of elements of the PMNS mixing matrix
in the low-energy theory. The dashed and solid lines are $\nu$ and $E$ indices, respectively.}
\end{figure}
the low-energy theory. Invariance under Eq.~(\ref{rph}) in the high-energy theory implies that every outgoing dashed or solid line must be connected to a corresponding incoming line.  Consequently, the connected graphs consist of closed loops with even numbers of vertices, Fig.~\ref{fig:loop},\ref{fig:quartic}, etc., and open chains beginning with an incoming $N^c$ line of a $W$ vertex and ending with an outgoing $N^c$ line emanating from a $W^*$ vertex, Fig.~\ref{fig:chain}, etc. Invariance under Eq.~(\ref{rpl}) in the low-energy effective theory implies that every outgoing solid line must be connected to an incoming solid line.  In this case, the connected graphs consist of 2-vertex chains beginning with an incoming $\nu$ line of a $U$ vertex and ending with an outgoing $\nu$ line of a $U^*$ vertex, Fig.~\ref{fig:4}.

The discrete $\eta_{\nu, N}$ invariance will be considered after we have constructed the basic loop and chain invariants. It requires that one consider only products of the basic chains where each external flavor index of a given type occurs an even number of times.

\subsection{Quark Invariants}
\label{subsec:quarkinvariants}

The classification of quark invariants has been studied in detail~\cite{greenberg,jarlskog,dunietz}. We review the known results as they will be needed for the discussion of lepton invariants. We also discuss relations among the invariants in a new way, which will help in understanding the structure of $CP$-violating phases in the lepton sector. We follow, to a large extent, the analysis of Nieves and Pal~\cite{np1}.

\begin{figure}
\begin{center}
\begin{pspicture}(0.5,0.5)(3.5,3.5)
\rput(2,1){\circlenode{D}{$\phantom{V}$}}
\rput(2,3){\circlenode{U}{$\phantom{V}$}}
\pnode(3,2){R}
\pnode(1,2){L}
\nccurve[linestyle=dashed,angleA=0,angleB=-90]{D}{R}
\nccurve[linestyle=dashed,angleA=90]{R}{U}
\nccurve[angleA=180,angleB=90]{U}{L}
\nccurve[angleA=-90,angleB=180]{L}{D}
\psline[arrowsize=0.2]{<-}(3,2.1)(3,1.9)
\psline[arrowsize=0.2]{->}(1,2.1)(1,1.9)
\rput(0.75,2){$\alpha$}
\rput(3.25,2){$i$}
\end{pspicture}
\end{center}
\caption{\label{fig:loop} Loop invariant $\langle i \alpha \rangle \equiv V_{i \alpha} V^*_{i \alpha}$.}
\end{figure}
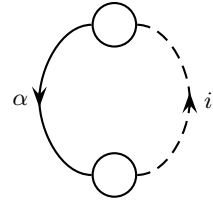

The quark invariants are constructed from the CKM matrix $V$, which has the same graphical depiction as the lepton mixing matrix $V$ of the seesaw theory shown in Fig.~\ref{fig:1}. Invariance under Eq.~(\ref{ckm}) implies that every outgoing dashed or solid line must be connected to a corresponding incoming line.  Thus, the rephasing invariants correspond graphically to closed loops involving an even number of  vertices.  Each invariant is the product of pairs of $V$ and $V^*$ matrix elements.  The intermediate lines in the loop graphs carry labels which specify the  $V$ and $V^*$ matrix elements in a given invariant.  The simplest closed loop involving two  vertices is displayed graphically in~Fig.~\ref{fig:loop}.  This loop denotes the rephasing invariant $V_{i \alpha} V_{i \alpha}^*$, where $V_{i \alpha}$ refers to the $i \alpha$ matrix element of $V$, and there is no implied summation over $i$ and $\alpha$.
A general rephasing invariant will be denoted by angle brackets $\loop{\ }$ surrounding the labels of the intermediate lines in the loop,
\begin{eqnarray}
\loop{ i_N \, \alpha_N \ldots i_2 \, \alpha_2 \, i_1 \, \alpha_1 }&\equiv&  V_{i_N \alpha_1}^* V_{i_N\alpha_N  } \ldots  V_{i_1 \alpha_2}^* V_{i_1 \alpha_1 }, \nn
\label{13}
\end{eqnarray}
where by convention, the first label inside the $\loop{\ }$ is always an $U$-quark $i$-type index. From the definition Eq.~(\ref{13}), it is clear that each loop rephasing invariant contains each $U$-quark label $i$ and $D$-quark label $\alpha$ twice on the r.h.s., once in a factor of $V$ and once in a factor of $V^*$.  It is implicit that each label takes one specific value in the set $1, \cdots, \ng$, and is not summed over, i.e.\ the Einstein summation convention is not implied.  It is clear that the loop invariant has cyclic symmetry,
\begin{eqnarray}
\loop{ i_N \, \alpha_N \ldots i_2 \, \alpha_2 \, i_1 \, \alpha_1 }
&=& \loop{i_1 \, \alpha_1 \, i_N \, \alpha_N \ldots i_2 \, \alpha_2 }. \nn
\end{eqnarray}
Under $CP$,
\begin{eqnarray}
\loop{ i_N \, \alpha_N \ldots i_2 \, \alpha_2 \, i_1 \, \alpha_1 }&\to&\loop{ i_N \, \alpha_N \ldots i_2 \, \alpha_2 \, i_1 \, \alpha_1 }^* \nn
&\equiv& \loop{ i_1 \, \alpha_2 \, i_2 \, \alpha_3 \ldots i_{N-1} \, \alpha_N \, i_N \, \alpha_1 }. \nn
\end{eqnarray}

Not all loop invariants are independent. The internal line labels of independent loop invariants must all take different values $1, \cdots, \ng$, because if any label of $U$-quark or $D$-quark type is repeated, the loop graph decomposes into smaller invariant subgraphs obtained by reconnecting the lines with the same labels.  For example, there are identities 
\begin{eqnarray}
\loop{[i \ldots \beta] \, j \alpha \, [\ell \ldots \gamma] \, k \alpha } &=& \loop{[i \ldots \beta]  \, j \alpha } \loop{ [\ell \ldots \gamma]\, k \alpha}\nn
\label{idone}
\end{eqnarray}
where $[i \ldots \beta]$ denotes an arbitrary string of allowed indices $i \cdots \beta$ with initial label $i$ and final label $\beta$, and $\alpha$ is a repeated $D$-quark label.  
This identity is the trivial equation
\begin{eqnarray}
&& V^*_{i \alpha} \cdots V^*_{j \beta} V_{j \alpha} V^*_{\ell \alpha} \cdots V^*_{k \gamma} V_{k \alpha} \nn
&=&  \left( V^*_{i \alpha} \cdots V^*_{j \beta} V_{j \alpha}\right) \left(  V^*_{\ell \alpha} \cdots V^*_{k \gamma} V_{k \alpha} \right) \ .
\end{eqnarray}
Similar identities hold for a repeated $U$-type index.  Thus, independent invariants have every quark label of a given type ($U$ or $D$) taking a distinct value.

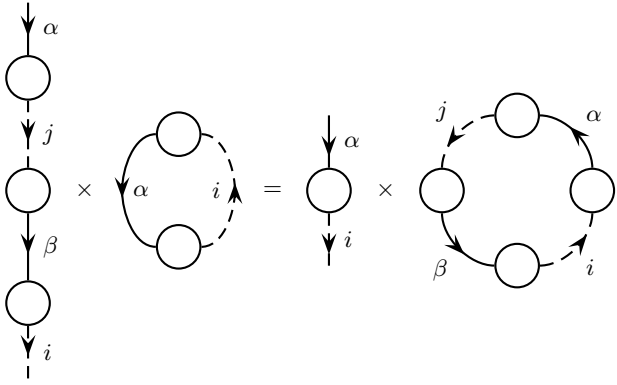
\begin{figure}
\begin{center}
\begin{pspicture}(-0.5,-0.5)(7.5,5)
\rput(2,1.25){\circlenode{D}{$\phantom{V}$}}
\rput(2,2.75){\circlenode{U}{$\phantom{V}$}}
\pnode(2.75,2){R}
\pnode(1.25,2){L}
\nccurve[linestyle=dashed,angleA=0,angleB=-90]{D}{R}
\nccurve[linestyle=dashed,angleA=90]{R}{U}
\nccurve[angleA=180,angleB=90]{U}{L}
\nccurve[angleA=-90,angleB=180]{L}{D}
\psline[arrowsize=0.2]{<-}(2.75,2.1)(2.75,1.9)
\psline[arrowsize=0.2]{->}(1.25,2.1)(1.25,1.9)
\rput(1.5,2){$\alpha$}
\rput(2.5,2){$i$}
\rput(0.75,2){$\times$}
\rput(0,0.5){\circlenode{v3}{$\phantom{V}$}}
\rput(0,2){\circlenode{v2}{$\phantom{V}$}}
\rput(0,3.5){\circlenode{v1}{$\phantom{V}$}}
\pnode(0,4.5){v0}
\pnode(0,-0.5){v4}
\ncline{v0}{v1}
\Aput{$\alpha$}
\ncline[linestyle=dashed]{v1}{v2}
\Aput{$j$}
\ncline{v2}{v3}
\Aput{$\beta$}
\ncline[linestyle=dashed]{v3}{v4}
\Aput{$i$}
\psline[arrowsize=0.2]{->}(0,4.15)(0,4.05)
\psline[arrowsize=0.2]{->}(0,2.75)(0,2.65)
\psline[arrowsize=0.2]{->}(0,1.25)(0,1.15)
\psline[arrowsize=0.2]{->}(0,-0.1)(0,-0.2)
\rput(3.25,2){$=$}
\pnode(4,3){w0}
\pnode(4,1){w2}
\rput(4,2){\circlenode{w1}{$\phantom{V}$}}
\ncline{w0}{w1}
\Aput{$\alpha$}
\ncline[linestyle=dashed]{w1}{w2}
\Aput{$i$}
\psline[arrowsize=0.2]{->}(4,2.55)(4,2.45)
\psline[arrowsize=0.2]{->}(4,1.25)(4,1.15)
\rput(6.5,1){\circlenode{D1}{$\phantom{V}$}}
\rput(6.5,3){\circlenode{U1}{$\phantom{V}$}}
\rput(7.5,2){\circlenode{R1}{$\phantom{V}$}}
\rput(5.5,2){\circlenode{L1}{$\phantom{V}$}}
\nccurve[linestyle=dashed,angleA=0,angleB=-90]{D1}{R1}
\Bput{$i$}
\nccurve[angleA=90]{R1}{U1}
\Bput{$\alpha$}
\nccurve[linestyle=dashed,angleA=180,angleB=90]{U1}{L1}
\Bput{$j$}
\nccurve[angleA=-90,angleB=180]{L1}{D1}
\Bput{$\beta$}
\psline[arrowsize=0.2]{->}(7.33,1.28)(7.43,1.41)
\psline[arrowsize=0.2]{->}(7.3,2.75)(7.2,2.85)
\psline[arrowsize=0.2]{->}(5.7,1.25)(5.8,1.15)
\psline[arrowsize=0.2]{->}(5.67,2.72)(5.57,2.59)
\rput(4.75,2){$\times$}
\end{pspicture}
\end{center}
\caption{\label{fig:iden} Identity which allows the 3 matrix elements $V V^* V$ to be replaced by a single matrix element $V$ in loop graphs.  Repeated application of the identity allows all loop graphs to be reduced down to 4-vertex and 2-vertex loop graphs.}
\end{figure}

There also are reconnection identities of the form
\begin{eqnarray}
&&\loop{{[i_1 \ldots \alpha_1] i \alpha [i_2 \ldots  \alpha_2] j \beta }}\loop{{[j_1 \ldots \beta_1] i^\prime \alpha [j _2 \ldots \beta_2]  j^\prime \beta }} \nn
&=& \loop{[i_1 \ldots \alpha_1] i \alpha [j _2\ldots \beta_2] j^\prime \beta }\loop{[j_1 \ldots\beta_1] i^\prime \alpha [i_2 \ldots \alpha_2] j \beta }\nn
\label{idtwo}
\end{eqnarray}
for repeated labels $\alpha$ and $\beta$. A particularly useful identity, obtained when one repeated label is a $U$-quark index and the other repeated label is a $D$-quark index, is
\begin{eqnarray}
\loop{[k \ldots \gamma] i \beta j \alpha } \loop{i \alpha}
&=& \loop{[k \ldots \gamma] i \alpha } \loop{i \beta j \alpha}
\label{24}
\end{eqnarray}
or 
\begin{eqnarray}
&&\left( V^*_{k \alpha} \cdots V^*_{i \gamma} V_{i \beta} V^*_{j \beta} V_{j \alpha} \right)
\left( V^*_{i \alpha} V_{i \alpha} \right) \nn
&=& \left( V^*_{k \alpha} \cdots V^*_{i \gamma} V_{i \alpha}\right) \left( V^*_{i \alpha} V_{i \beta} V^*_{j \beta} V_{j \alpha} \right) \ .
\end{eqnarray}
This identity replaces three matrix elements $V V^* V$  in the loop $\loop{[k \ldots \gamma] i \beta j \alpha }$ by one matrix element $V$ (see Fig.~\ref{fig:iden}). By repeatedly applying Eq.~(\ref{24}), any loop can be reduced down to products of loops containing at most 4 vertices.

First, we determine the independent quadratic $V V^*$ invariants.
Unitarity of the CKM matrix,
\begin{eqnarray}
\sum_i V_{i \alpha} V_{i \beta}^* &=& \delta_{\alpha \beta},\nn
\sum_\alpha V_{i \alpha}^* V_{j \alpha} &=& \delta_{ij},
\label{uni}
\end{eqnarray}
which is depicted graphically in Fig.~\ref{fig:7}, implies that all invariants with an internal index equal to $\ng$ can be written in terms of the other quadratic invariants.  Thus, the independent quadratic invariants are given by $\loop{i \alpha}$ with $i$ and $\alpha$ labels running over $1,2, \cdots, \ng-1$,
\begin{eqnarray}
\loop{i \alpha} &\equiv& \abs{V_{i \alpha}}^2,\qquad  1 \le i,\alpha \le \ng-1,
\end{eqnarray}
and they are all $CP$-even. There are $(\ng-1)^2$ independent quadratic invariants $\loop{i \alpha}$, which is equal to the number of parameters (angles plus phases) in $V$.
\emph{These quadratic invariants generically determine all the parameters of $V$ except for discrete choices.} For example, for $\ng=3$ they determine $\theta_{1,2,3}$ and $\cos \delta$, which determines $\delta$ up to a $\pm$ sign.
\begin{figure}
\begin{center}
\begin{pspicture}(-1,0.5)(8,1.5)
\rput(1.75,1){\circlenode{L}{$\phantom{V}$}}
\rput(3.25,1){\circlenode{R}{$\phantom{V}$}}
\rput(0.5,1){\rnode{l}{$i\ $}}
\rput(4.5,1){\rnode{r}{$\ j$}}
\ncline[linestyle=dashed]{r}{R}
\ncline[linestyle=dashed]{L}{l}
\ncline{R}{L}
\Bput{$\alpha$}
\psline[arrowsize=0.2]{<-}(0.85,1)(0.9,1)
\psline[arrowsize=0.2]{<-}(2.4,1)(2.5,1)
\psline[arrowsize=0.2]{<-}(3.85,1)(3.9,1)
\rput(0,0.9){$\displaystyle\sum_\alpha$}
\rput(5.25,1){$=$}
\rput(6,1){\rnode{i2}{$i\ $}}
\rput(7,1){\rnode{j2}{$\ j$}}
\ncline[linestyle=dashed]{j2}{i2}
\psline[arrowsize=0.2]{<-}(6.4,1)(6.5,1)
\rput(7.5,1){$\delta_{ij}$}
\end{pspicture}
\end{center}
\begin{center}
\begin{pspicture}(-1,0.5)(8,1.5)
\rput(1.75,1){\circlenode{L}{$\phantom{V}$}}
\rput(3.25,1){\circlenode{R}{$\phantom{V}$}}
\rput(0.5,1){\rnode{l}{$\alpha\ $}}
\rput(4.5,1){\rnode{r}{$\ \beta$}}
\ncline{r}{R}
\ncline{L}{l}
\ncline[linestyle=dashed]{R}{L}
\Bput{$i$}
\psline[arrowsize=0.2]{<-}(0.85,1)(0.9,1)
\psline[arrowsize=0.2]{<-}(2.4,1)(2.5,1)
\psline[arrowsize=0.2]{<-}(3.85,1)(3.9,1)
\rput(0,0.9){$\displaystyle\sum_i$}
\rput(5.25,1){$=$}
\rput(6,1){\rnode{i2}{$\alpha\ $}}
\rput(7,1){\rnode{j2}{$\ \beta$}}
\ncline{j2}{i2}
\psline[arrowsize=0.2]{<-}(6.4,1)(6.5,1)
\rput(7.6,1){$\delta_{\alpha\beta}$}
\end{pspicture}
\end{center}
\caption{\label{fig:7} Graphical representation of unitarity of the matrix $V_{i \alpha}$.}
\end{figure}

\begin{figure}
\begin{center}
\begin{pspicture}(0.5,0.5)(3.5,3.5)
\rput(2,1){\circlenode{D}{$\phantom{V}$}}
\rput(2,3){\circlenode{U}{$\phantom{V}$}}
\rput(3,2){\circlenode{R}{$\phantom{V}$}}
\rput(1,2){\circlenode{L}{$\phantom{V}$}}
\nccurve[linestyle=dashed,angleA=0,angleB=-90]{D}{R}
\Bput{$i$}
\nccurve[angleA=90]{R}{U}
\Bput{$\alpha$}
\nccurve[linestyle=dashed,angleA=180,angleB=90]{U}{L}
\Bput{$j$}
\nccurve[angleA=-90,angleB=180]{L}{D}
\Bput{$\beta$}
\psline[arrowsize=0.2]{->}(2.83,1.28)(2.93,1.41)
\psline[arrowsize=0.2]{->}(2.8,2.75)(2.7,2.85)
\psline[arrowsize=0.2]{->}(1.2,1.25)(1.3,1.15)
\psline[arrowsize=0.2]{->}(1.17,2.72)(1.07,2.59)
\end{pspicture}
\end{center}
\caption{\label{fig:quartic} Loop invariant $\langle i \alpha j \beta \rangle = V_{i \alpha} V^*_{j \alpha} V_{j \beta} V^*_{i \beta} $.}
\end{figure}
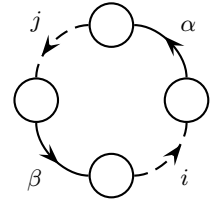

Next, we consider the quartic invariants (see Fig.~\ref{fig:quartic})
\begin{eqnarray}
\loop{i \alpha j \beta} &\equiv& V_{i \alpha} V_{j \alpha}^* V_{j \beta} V_{i \beta}^* .
\end{eqnarray}
Not all of these quartic invariants are independent.  The identities for the quartic invariants are
\begin{eqnarray}
\loop{i \alpha \, j \alpha }&=& \loop{i \alpha }\loop{j \alpha },\nn
\loop{i \alpha \, i \beta }&=&\loop{i \alpha }\loop{i \beta },\nn
\loop{i \alpha j \beta }^* \loop{i \alpha j \beta } &=& \loop{i \alpha } \loop{i \beta }
\loop{j \alpha }\loop{j \beta },\nn
\loop{i \alpha \, j \beta } \loop{k \alpha \, \ell \beta } &=& \loop{i \alpha \, \ell \beta} \loop{k \alpha \, j \beta }, \nn
\loop{i \alpha \, j \beta } \loop{i \gamma \, j \delta } &=& \loop{i \alpha \, j \delta }
\loop{i \gamma \, j \beta },
\label{4id}
\end{eqnarray}
where the first two identities follow from Eq.~(\ref{idone}), the last two identities follow from Eq.~(\ref{idtwo}), and the third identity is the trivial statement that the absolute value of the magnitude squared of any loop invariant can be written as the product of quadratic loop invariants.  The first two identities restrict the quartic invariants $\loop{i \alpha j \beta}$ to ones with no repeated labels of the same type, $i \not= j$ and $\alpha \not= \beta$.  Unitarity Eq.~(\ref{uni}) implies that the quartic loop invariants with $i$ and $\alpha$ set to $\ng$ can be rewritten in terms of invariants with $i$ and $\alpha$ equal to $1, \cdots, \ng -1$.     
The identities in Eq.~(\ref{4id}) also yield
\begin{eqnarray}
\loop{i \alpha j \beta } \loop{j \alpha k \beta} &=& \loop{i \alpha k \beta } \loop{j \alpha }
\loop{j \beta }, \nn
\loop{i \alpha j \beta } \loop{i \beta j \gamma } &=& \loop{i \alpha j \gamma }
\loop{i \beta }\loop{j \beta}.
\end{eqnarray}
These relations can be used to express~\cite{np1}
\begin{eqnarray}
\loop{i \alpha j \beta } &=& \frac{\loop{i \alpha i_0 \alpha_0 } \loop{j \beta i_0 \alpha_0}
\loop{j \alpha i_0 \alpha_0 }^* \loop{i \beta i_0 \alpha_0 }^*}{
\loop{i_0 \alpha_0 }^2 \loop{i_0 \alpha } \loop{i_0 \beta }\loop{i \alpha_0}\loop{j \alpha_0 }}\nn
\label{22}
\end{eqnarray}
in terms of quartic loop invariants with the last two labels set equal to the reference values $i_0 \alpha_0$.  Choosing the fixed labels $i_0$ and $\alpha_0$ to be $1$ in Eq.~(\ref{22}), one sees that all of the $\loop{i \alpha j \beta}$ can be obtained in terms of the quartic invariants $\loop{k \gamma i_0 \alpha_0}$, with $2 \le k, \gamma \le \ng-1$. The index value $\ng$ is eliminated by unitarity, and the index $1$ is eliminated because repeated indices are not allowed, $k \not= i_0 =1$ and $\gamma \not= \alpha_0 =1$.

The analysis so far shows that all invariants can be written in terms of $\loop{i \alpha}$,  $1 \le i, \alpha \le \ng-1$ and $\loop{j \beta i_0 \alpha_0}$, $2 \le j, \beta \le \ng-1$, $i_0 = \alpha_0 =1$.  The $\loop{i \alpha}$ are $(\ng-1)^2$ $CP$-even invariants, and the $\loop{j \beta i_0 \alpha_0}$ are $(\ng-2)^2$ complex invariants, with a $CP$-even real part and a $CP$-odd imaginary part.  There are further nonlinear relations among the remaining invariants, that follow from
\begin{eqnarray}
T_{i \alpha} = \frac{\loop{i \alpha i_0 \alpha_0}}{\sqrt{ \loop{i_0 \alpha_0 } \loop{i_0 \alpha } \loop{i \alpha_0}}}
\label{tmatrix}
\end{eqnarray}
being a unitary matrix, as was pointed out by Nieves and Pal~\cite{np1}.

It turns out that $T_{i\alpha}$ is essentially the original CKM matrix. Writing out the loop invariants explicitly, one finds
\begin{eqnarray}
T_{i \alpha} &=& V_{i \alpha} e^{i \phi(V_{i_0 \alpha_0})-i\phi(V_{i_0 \alpha})-i\phi(V_{i \alpha_0})}
\end{eqnarray}
where $\phi(V_{i\alpha})$ is the phase of the matrix element $V_{i \alpha}$.
If one picks the standard form $\mathcal{V}$ such that the $i_0$ row and $\alpha_0$ column are real and non-negative, then $T_{i \alpha}$ is \emph{identical} to the original CKM matrix $V$! For other forms, $T$ is the matrix $V$ with phase rotations on the rows and columns to make the $i_0$ row and $\alpha_0$ column real and non-negative.

It seems that we have ended up with a circular analysis, characterizing $V$ in terms of $T$, which is, in fact, identical to $V$. This is not the case. The quadratic invariants $\loop{i \alpha}$ already determine $\abs{V_{i \alpha}}$, so there is only a discrete amount of information in the quartic invariants $T_{i\alpha}$. This result is true for any number of generations, but it is best explained by considering the special cases $\ng=2,3$ before discussing general $\ng$.

\subsubsection{$\ng=2$}

The only independent quadratic invariant is $\loop{11}$, and there are no independent quartic invariants. The quadratic invariant $\loop{11}\equiv  |V_{11}|^2 =\cos^2\theta_C$ determines the Cabibbo angle $\theta_C$, where $\theta_C \in [0, \pi/2]$ is restricted to the first quadrant. All elements of $T_{i \alpha}$ are determined in terms of $\loop{11}$.

\subsubsection{$\ng=3$}
\label{subsec:quark3}

There are four independent quadratic invariants 
\begin{eqnarray}
\loop{11} &\equiv& |V_{11}|^2, \nn
\loop{12} &\equiv& |V_{12}|^2, \nn 
\loop{21} &\equiv& |V_{21}|^2, \nn 
\loop{22} &\equiv& |V_{22}|^2, 
\end{eqnarray}
which are all $CP$-even.  The five quadratic invariants $\loop{13}$, $\loop{23}$, $\loop{33}$, $\loop{32}$ and $\loop{31}$ are determined in terms of these four using the unitarity relations for $V$.  The number of independent quadratic invariants is equal to the total number of parameters (angles and phases) of the CKM matrix.  These four quadratic invariants can be used to determine the four parameters $\cos \theta_i$, $i=1,2,3$ and $\cos \delta$.  Since $\theta_i \in [0, \pi/2]$, the individual angles are determined from knowledge of the $\cos \theta_i$.  However, $\delta \in [0, 2\pi)$ is not determined unambiguously from the value of $\cos \delta$; $\delta$ is determined only up to a two-fold ambiguity. Even though $\delta$ itself is $CP$-odd, $\cos \delta$ is $CP$-even and is fixed by $CP$-even invariants. The remaining piece of information needed to determine $\delta$ is a $\mathbb{Z}_2$ factor which is $CP$-odd, the \emph{sign} of $\sin \delta$. The $CP$-odd quantity determining the sign of $\delta$ is the only remaining information contained in the quartic invariants.\footnote{ This discussion  is related to the well-known result that one can determine that the unitarity triangle has non-zero area, which is a $CP$-odd quantity, by measuring the lengths of its sides, which are $CP$-even quantities.  There still remains a two-fold ambiguity between the triangle and its mirror image (i.e.\ the sign of the area), which is resolved by measuring a $CP$-odd quantity.}

For $\ng =3$, there is only one independent quartic invariant, $\loop{2211}$. The standard Jarlskog invariant is the imaginary part of this invariant,
\begin{eqnarray}
J \equiv \text{Im}\, \loop{2211} = \text{Im} \, V_{11} V_{12}^* V_{22} V_{21}^* .
\end{eqnarray}  
There are non-trivial relations between the quartic invariant $\loop{2211}$ and the four independent quadratic invariants. The real part of $\loop{2211}$, which is $CP$-even, is determined by the quadratic invariants,
\begin{eqnarray}
\loop{2211} +\loop{2211}^* &=& 1-\Bigl[ \loop{11}+\loop{22} + \loop{12} + \loop{21} \Bigr]\nn
&& + \loop{11} \loop{22} +\loop{12}\loop{21}
\label{q1}
\end{eqnarray}
as is the absolute magnitude squared of $\loop{2211}$, which also is $CP$-even, 
\begin{eqnarray}
\loop{2211} \loop{2211}^* &=&  \loop{11} \loop{22} \loop{12}\loop{21} \ .
\label{q2}
\end{eqnarray}
Thus, both ${\rm Re}\loop{2211}$ and $J^2 = [{\rm Im}\,\loop{2211}]^2$ are determined by the quadratic invariants.  The only new piece of information in the quartic invariant $\loop{2211}$ is the sign of the Jarlskog invariant $J ={\rm Im}\,\loop{2211}$. 

\subsubsection{$\ng \ge 4$}

The $(\ng-1)^2$ quadratic invariants $\loop{i \alpha}$, $1 \le i,\alpha \le \ng-1$ determine $\abs{V_{i \alpha}}^2$.  The total number of angles and phases is equal to $(\ng -1)^2$, so the number of independent quadratic invariants is equal to the total number of parameters.  The quadratic invariants determine the CKM matrix up to discrete ambiguities. Thus, we conclude that the only remaining information contained in the quartic invariants $T_{i \alpha}$ is discrete information about $V_{i \alpha}$. This observation can be made more precise by posing the following mathematical question, to which we do not know the general answer: 

\medskip

\noindent \emph{If $V$ is  a $\ng \times \ng$ unitary matrix, what are the  allowed $\ng \times \ng$ unitary matrices $T$ such that the corresponding elements of $V$ and $T$ have the same magnitude, i.e.\ $\abs{V_{i\alpha}}=\abs{T_{i \alpha}}$?}

\medskip

\noindent 
We will refer to such matrices as isomodular unitary matrices.
There are trivial phase redefinitions of $T$ given by multiplying it on the left and right by a diagonal unitary matrix. To eliminate these, $V$ and $T$ are restricted so that the $i_0$ row and $\alpha_0$ column are real and non-negative.  [By setting $i_0=\alpha_0=1$, both matrices have the first row and column real and non-negative.] This eliminates all the phase redefinitions in the generic case where all entries in the $i_0$ row and $\alpha_0$ column are non-zero.
 
For $\ng=2$, the only solution is $T=V$.  For $\ng=3$, there are two solutions $T=V$ and $T=V^*$, which have opposite signs for the Jarlskog invariant. For $\ng \ge 4$, there are other solutions in addition to $V$ and $V^*$, which are distinguished by the values of their quartic invariants, but we have been unable to classify them in full generality. Generically, there will be an even number of solutions, since if $T$ is a solution, so is $T^*$.  Note that a simple sign change of one of the phases, e.g. $\delta_1 \to - \delta_1$ for $\ng=4$ does not lead to a solution.

The problem of determining the isomodular unitary matrices for $\ng=4$ has been studied before. It has been shown that generically there are eight discrete solutions (4 plus their complex conjugates)~\cite{lavoura}. For certain special values of the mixing angles, there is a continuous family of solutions~\cite{auberson}.

Nieves and Pal~\cite{np1} use the $(\ng-1)(\ng-2)/2$ invariants $\text{Im}\, T_{i \alpha}$ (or equivalently $\text{Im}\, \loop{i \alpha 11}$) for $2 \le i \le \alpha \le \ng -1$ to fix $V$. These are sufficient to uniquely fix $V$, since they determine all the CKM phases $\delta_i$. However, most of the information in $\text{Im}\, T_{i \alpha}$ has already been determined by the quadratic invariants, as shown in Eqs.~(\ref{q1}) and~(\ref{q2}). In what follows, we will use this redundant choice, since we do not know the general solution to the question posed above.

\subsection{PMNS invariants}
\label{subsec:pmns}

In the lepton sector, the rephasing invariants of the low-energy theory are constructed from the PMNS matrix elements shown in Fig.~\ref{fig:2}.  Invariance under the $\Phi_E$ rephasings of Eq.~(\ref{rpl}) implies that the invariants correspond to graphs with no external solid lines, such as the 2-vertex chains shown in Fig.~\ref{fig:4}.
The basic quadratic invariants will be denoted by
\begin{eqnarray}
\uchain{j \alpha i} &\equiv& U_{\alpha i} U_{\alpha j}^*, 
\end{eqnarray}
and were considered previously by Nieves and Pal~\cite{np1,np2}. Under $CP$,
\begin{eqnarray}
\uchain{j \alpha i} &{\longrightarrow}& \uchain{j\alpha i}^*= \uchain{i \alpha j}.
\end{eqnarray}
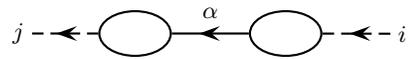
\begin{figure}
\begin{center}
\begin{pspicture}(0,0.5)(5,1.5)
\rput(1.5,1){\ovalnode{U1}{$\phantom{U_{\alpha i}}$}}
\rput(3.5,1){\ovalnode{U2}{$\phantom{U_{\alpha i}}$}}
\rput(0,1){\rnode{i}{$j\ $}}
\rput(5,1){\rnode{j}{$\ i$}}
\ncline[linestyle=dashed]{i}{U1}
\ncline{U1}{U2}
\Aput{$\alpha$}
\ncline[linestyle=dashed]{U2}{j}
\psline[arrowsize=0.2]{<-}(2.35,1)(2.45,1)
\psline[arrowsize=0.2]{<-}(0.45,1)(0.55,1)
\psline[arrowsize=0.2]{<-}(4.35,1)(4.45,1)
\end{pspicture}
\end{center}
\caption{\label{fig:4} Chain invariant $\uchain{j \alpha i} \equiv U_{\alpha i} U^*_{\alpha j}$  constructed from the PMNS matrix. The dashed and solid lines are $\nu$ and $E$ indices, respectively.}
\end{figure}

The PMNS invariants must also be invariant under the discrete $\eta_\nu$ tranformation in Eq.~(\ref{rpl}), which was not considered previously. Under this discrete symmetry, $\uchain{j \alpha i} \to \uchain{j \alpha i} (\eta_\nu)_{jj} (\eta_\nu)_{ii}$, and is not invariant unless $j =i$.  Quartic invariants which are $\eta_\nu$ invariant are products of the basic quadratic $\uchain{j \alpha i}$ in which each external index occurs an even number of times, since the $\uchain{ j \alpha i}$, $j \not=i$, individually are not invariant.
The quartic invariants are $\uchain{j \alpha i}^2$ and $\uchain{j \alpha i} \uchain{i \alpha j} = |\uchain{j \alpha i}|^2$ for $j \not= i$, and $\uchain{j \alpha i} \uchain{j \beta i}$ and $\uchain{j \alpha i}\uchain{i \beta j}$ for $j \not= i$ and $\alpha \not= \beta$.  The $\eta_\nu$-invariant quantities depend on $\Psi$ rather than $\Psi/2$.

Not all of these invariants are independent.  Unitarity of the PMNS matrix implies that
\begin{eqnarray}
\sum_\alpha U_{\alpha i} U^*_{\alpha j} &=& \delta_{ij}, \nn
\sum_i U_{\alpha i} U^*_{\beta i} &=& \delta_{\alpha \beta},
\label{pmnsunitarity}
\end{eqnarray}
which yields the identities
\begin{eqnarray}
\sum_\alpha \uchain{j \alpha i } &=& \delta_{ij},\nn
\sum_i \uchain{ i \alpha i} &=& 1,
\label{uid1}
\end{eqnarray}
where the second set of identities correspond to only the $\ng$ diagonal equations of the second unitarity constraint in Eq.~(\ref{pmnsunitarity}).  The identities Eq.~(\ref{uid1}) can be used to eliminate $(2\ng -1)$ quadratic invariants; we choose to eliminate the $\ng$ invariants $\uchain{ \ng \alpha \ng}$, $1 \le \alpha \le \ng$, and the $(\ng -1)$ invariants $\uchain{i \ng i}$, $1 \le i \le \ng-1$.  With this choice, the independent quadratic invariants are
\begin{eqnarray}
\uchain{i \alpha i} &\equiv& | U_{\alpha i}|^2,\ 1 \le \alpha, i \le \ng -1 .
\end{eqnarray} 
These $(\ng -1)^2$ quadratic invariants are all $CP$-even; they determine the $\mathcal{V}$ part of the PMNS matrix Eq.~(\ref{Ustd}) up to discrete ambiguities, as for the CKM case discussed in the previous section, and are the analogs of $\loop{i \alpha}$.

There are many identities which can be used to eliminate most of the quartic invariants,
such as
\begin{eqnarray}
\uchain{ j \alpha i} \uchain{ l \alpha k } &=& \uchain{ j \alpha k} \uchain{ l \alpha i }.
\label{quad}
\end{eqnarray}
Setting $k=l=i_0$ gives
\begin{eqnarray}
\uchain{ j \alpha i }\uchain{ i_0 \alpha i_0 } &=& \uchain{ j \alpha i_0 }\uchain{i_0 \alpha i}=  \uchain{i_0 \alpha j}^* \uchain{i_0 \alpha i}
\label{30}
\end{eqnarray}
so that $\uchain{j\alpha i}$ is determined by $\uchain{i_0 \alpha i}$, $\uchain{i_0 \alpha j}$ and $\uchain{i_0 \alpha i_0}$,
\begin{eqnarray}
\uchain{ j \alpha i} &=& { {\uchain{ j \alpha i_0} \uchain{i_0 \alpha i}} \over \uchain{i_0 \alpha i_0}} .
\label{42}
\end{eqnarray}
The relation
\begin{eqnarray}
\uchain{j \alpha i}\uchain{i \alpha j} &=& |\uchain{j \alpha i}|^2 = \uchain{i \alpha i} \uchain{j \alpha j},
\label{43} 
\end{eqnarray}
implies that all $|\uchain{j \alpha i}|^2$ can be determined from the quadratic invariants $\uchain{i \alpha i}$. There are also identities for invariants $\uchain{ j \alpha i} \uchain{i \beta j}$, $\alpha \not = \beta$ and $j \not=i$, such as
\begin{eqnarray}
\sum_i \uchain{ j \alpha i} \uchain{ i \beta j} &=& 
\uchain{ j \alpha j}\ \delta_{\alpha \beta},\
\end{eqnarray}
which is $\eta_\nu$-invariant, and follows from unitarity of $U$.

Equations~(\ref{42},\ref{43}) show that $(i \alpha j)$ can be written in terms of the quadratic invariants $(i \alpha i)$, and $(i_0 \alpha i)$, for a fixed value $i_0$. For $\eta_\nu$ invariance, the $(i_0 \alpha i)$ factors must occur in pairs of the form $(i_0 \alpha i)(i_0 \beta i)$, so that $i$ occurs twice. It is straightforward to show from the above relations that all such combinations can be written in terms of $(i_0 \alpha_0 i)^2$ and $(i_0 \alpha i)(i \alpha_0 i_0)$ for a fixed value $\alpha_0$, the latter being the analog of $\loop{i \alpha i_0 \alpha_0}$ of Sec.~\ref{subsec:quarkinvariants}.

One can define a matrix $S$,
\begin{eqnarray}
S_{\alpha i} &=& \frac{ \uchain{i_0\alpha i }\uchain{i \alpha_0 i_0}}{\sqrt{ \uchain{i_0 \alpha_0 i_0}\uchain{i_0 \alpha  i_0}\uchain{i \alpha_0 i}}}
\label{smatrix}
\end{eqnarray}
in analogy to the $T$-matrix for the quark sector defined in Eq.~(\ref{tmatrix}).\footnote{The definition in Ref.~\cite{np1},
$S_{\alpha i} ={ \uchain{i_0\alpha i }}/{\sqrt{ \uchain{i_0 \alpha  i_0}}}$,
is not $\eta_\nu$ invariant.}
$S$ is a unitary matrix constructed out of rephasing invariants, and is equal to
\begin{eqnarray}
S_{\alpha i} &=& U_{\alpha i}\ e^{-i \phi(U_{\alpha i_0}) + i \phi(U_{\alpha_0 i_0}) -i \phi(U_{\alpha_0 i})}.
\label{smatrix2}
\end{eqnarray}
$S$ is \emph{identical} to the $\mathcal{V}$ matrix in the PMNS form of $U$ in Eq.~(\ref{Ustd}), where $\mathcal{V}$ is chosen so that row $\alpha_0$ and column $i_0$ are real and non-negative.

 $S$ is a unitary matrix, which gives the identities necessary to eliminate all of the dependent invariants. The information contained in $S$ is discrete information about $\mathcal{V}$. As for $T_{i\alpha}$, we can choose the redundant variables $\text{Im}\, S_{\alpha i}$, $2 \le \alpha \le i \le \ng-1$, which determine the phases $\delta_i$ in $\mathcal{V}$.

The independent quartic invariants determine the $(\ng -1)$ phases $\psi_i$ and fix the discrete ambiguity in $\mathcal{V}$. Consequently, only $CP$-odd quartic invariants are independent.  Without loss of generality, we choose $i_0 =1$, $\alpha_0 =1$.  With this choice, the independent quartic invariants are the $CP$-odd invariants
\begin{eqnarray}
&&{\rm Im}\, \uchain{1 \alpha i}\uchain{i 1 1},\ 2 \le \alpha \le  i \le \ng -1\nn[5pt]
&&{\rm Im}\, \uchain{ 11i}^2,\  \  2 \le i \le \ng.
\end{eqnarray}
The first set of $CP$-odd invariants determine the $(\ng -1)(\ng -2)/2$ phases $\delta_i$ in $\mathcal{V}$.  The second set of $CP$-odd invariants determine the $(\ng -1)$ phases $\psi_i$.

\subsubsection{$\ng=2$}
\label{subsec:pmns2}

The independent invariants are
\begin{eqnarray}
\uchain{111} &\equiv& | U_{11} |^2, \nn
{\rm Im}\, \uchain{ 112}^2 &\equiv& {\rm Im}\, \left( U_{12} U^*_{11} \right)^2.
\end{eqnarray}
The $2 \times 2$ matrix $U$ is parametrized by one angle and one phase.  The $CP$-even invariant $\uchain{ 111} = \cos^2 \theta^{(U)}$ determines the single angle $\theta^{(U)}$.  The $CP$-odd invariant ${\rm Im}\, \uchain{ 112}^2$ determines the single phase $\psi^{(U)}$.

\subsubsection{$\ng=3$}
\label{subsec:pmns3}

The independent quadratic invariants are the $CP$-even quantities
\begin{eqnarray}
\uchain{111} &\equiv& | U_{11} |^2, \nn
\uchain{121} &\equiv& | U_{21} |^2, \nn
\uchain{212} &\equiv& | U_{12} |^2, \nn
\uchain{222} &\equiv& | U_{22} |^2.
\end{eqnarray} 
The invariants $\uchain{111}$, $\uchain{121}$ and $\uchain{212}$ determine the three angles $\theta^{(U)}_{1,2,3}$, respectively.  Invariant $\uchain{222}$ determines $\cos \delta^{(U)}$, which gives $\delta^{(U)}$ up to a two-fold sign ambiguity.

The independent quartic invariants are the $CP$-odd quantities
\begin{eqnarray}
{\rm Im}\, \uchain{ 122} \uchain{211} &=& {\rm Im}\, U_{12}^* U_{11} U_{21}^* U_{22} , \nn
{\rm Im}\, \uchain{112}^2 &=& {\rm Im}\, \left( U_{12} U^*_{11} \right)^2 ,\nn
{\rm Im}\, \uchain{ 113}^2 &=& {\rm Im}\, \left( U_{13} U^*_{11} \right)^2.
\end{eqnarray}
The first invariant is  the analogue of the Jarlskog invariant for the PMNS matrix 
$U$. It gives the sign of $\delta^{(U)}$.  The second and third invariants determine $\psi^{(U)}_{2,3}$, respectively.

\subsubsection{$\ng \ge 4$}
\label{subsec:pmns4}

There are $(\ng -1)^2$ independent quadratic invariants
\begin{eqnarray}
\uchain{ i \alpha i} &=& | U_{\alpha i} |^2, \ 1 \le i, \alpha \le \ng -1.
\end{eqnarray} 
which determine the magnitudes of $U_{\alpha i}$, and a total of $\ng(\ng-1)/2$ independent quartic invariants consisting of
\begin{eqnarray}
&&{\rm Im}\, \uchain{11i}^2 = {\rm Im}\, \left( U_{1 i} U^*_{11} \right)^2,\  2 \le i \le \ng.
\end{eqnarray}
which determine the $\ng-1$ phases $\psi_i$, and 
\begin{eqnarray}
&& {\rm Im}\, \uchain{1 \alpha i }\uchain{i 11 },\ 2 \le  \alpha \le i \le \ng-1
\end{eqnarray}
which determine the $(\ng-1)(\ng-2)/2$ phases $\delta_i$ and fix the discrete ambiguity in $\mathcal{V}$.

\subsection{High Energy Invariants}

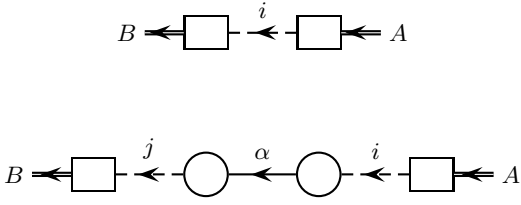
\begin{figure}
\begin{center}
\begin{pspicture}(0.5,0)(4,1.5)
\rput(1.5,1){\rnode{U1}{\framebox{$\phantom{W}$}}}
\rput(3,1){\rnode{U2}{\framebox{$\phantom{W}$}}}
\rput(0.5,1){\rnode{i}{$B\ $}}
\rput(4,1){\rnode{j}{$\ A$}}
\ncline[doubleline=true,doublesep=0.5\pslinewidth]{i}{U1}
\ncline[linestyle=dashed]{U1}{U2}
\Aput{$i$}
\ncline[doubleline=true,doublesep=0.5\pslinewidth]{U2}{j}
\psline[arrowsize=0.2]{<-}(2.15,1)(2.2,1)
\psline[arrowsize=0.2]{<-}(0.75,1)(0.8,1)
\psline[arrowsize=0.2]{<-}(3.35,1)(3.4,1)
\end{pspicture}
\end{center}
\begin{center}
\begin{pspicture}(0,0)(6.5,1.5)
\rput(1,1){\rnode{1}{\framebox{$\phantom{W}$}}}
\rput(5.5,1){\rnode{5}{\framebox{$\phantom{W}$}}}
\rput(2.5,1){\circlenode{2}{$\phantom{V}$}}
\rput(4,1){\circlenode{4}{$\phantom{V}$}}
\rput(0,1){\rnode{0}{$B\ $}}
\rput(6.5,1){\rnode{6}{$\ A$}}
\ncline[doubleline=true,doublesep=0.5\pslinewidth]{0}{1}
\ncline[linestyle=dashed]{1}{2}
\Aput{$j$}
\ncline{2}{4}
\Aput{$\alpha$}
\ncline[linestyle=dashed]{4}{5}
\Aput{$i$}
\ncline[doubleline=true,doublesep=0.5\pslinewidth]{5}{6}
\psline[arrowsize=0.2]{<-}(0.3,1)(0.35,1)
\psline[arrowsize=0.2]{<-}(1.6,1)(1.65,1)
\psline[arrowsize=0.2]{<-}(3.1,1)(3.25,1)
\psline[arrowsize=0.2]{<-}(4.6,1)(4.65,1)
\psline[arrowsize=0.2]{<-}(5.9,1)(5.95,1)
\end{pspicture}
\end{center}
\caption{\label{fig:chain} Chain invariants $\chain{BiA}\equiv W_{iA} W_{iB}^*$ and $\chain{ B j \alpha i A } \equiv W_{iA} V^*_{i \alpha} V_{j \alpha} W^*_{j B}$. }
\end{figure}

The invariants in the high-energy theory are 
made out of the $V$ and $W$ vertices shown in Fig.~\ref{fig:1}. It is easy to see that there are two forms for the invariants: closed loops involving an even number of $V$ matrices such as~Figs.~\ref{fig:loop} and \ref{fig:quartic}, and chains with an even number of $V$ vertices terminated by $W$ vertices at each end such as~Fig.~\ref{fig:chain}.

The loop invariants are denoted by $\loop{\ }$, and have the same form as the quark invariants Eq.~(\ref{13}) with $V_{i \alpha}$ now representing the lepton mixing matrix with $\nu$ and $E$ labels $i$ and $\alpha$, respectively.  The loop invariants involve only $V$ and so determine all of the parameters in the canonical CKM form of $V$ in Eq.~(\ref{vwone}). 

The chain invariants are denoted by $\chain{ \ }$, where
\begin{eqnarray}
\chain{ B j \beta \ldots  \alpha i A }&=& W_{i A} V^*_{i \alpha } \ldots  V_{j \beta } W^*_{j B}\nn
\end{eqnarray}
Under $CP$,
\begin{eqnarray}
\chain{B j \beta \ldots  \alpha i A  }&\to&\chain{ B j \beta \ldots  \alpha i A  }^* \nn
&&=\chain{A i \alpha \ldots \beta j B } .
\end{eqnarray}

The basic quadratic chain invariants
\begin{eqnarray}
\chain{B i A } &\equiv& W_{i A} W^*_{i B} 
\end{eqnarray}
involve only $W$ matrix elements, and are the analogues of the invariants $(j \alpha i)$ discussed in Sec.~\ref{subsec:pmns} for the PMNS matrix.  The discussion of independent PMNS invariants applies to these new invariants $\chain{B i A}$, which determine all of the parameters in the canonical PMNS form of $W$ in Eq.~(\ref{vwtwo}). 

There remain the $\ng-1$ phases in $\bar \Phi$ which can be included in either $V$ or $W$, Eqs.~(\ref{vwone}) and~(\ref{vwtwo}). These phases are determined by the chain invariants
\begin{eqnarray} 
\chain{A j \alpha i A } &\equiv& W_{i A} V^*_{i \alpha} V_{j \alpha} W^*_{j A} 
\end{eqnarray} 
which involve both $V$ and $W$ matrix elements, and are invariant under the full rephasing transformation Eq.~(\ref{rph}).  The independent invariants, as shown below, are the $\ng -1$ $CP$-odd invariants
\begin{eqnarray}
{\rm Im}\, \chain{ A_0 i_0 \alpha_0 i A_0} &\equiv& {\rm Im}\, W_{i A_0} V^*_{i \alpha_0} V_{i_0 \alpha_0} W^*_{i_0 A_0} , \ i \not= i_0 ,\nn
\label{newchain}
\end{eqnarray}
for reference values $A_0$, $i_0$ and $\alpha_0$.  Choosing the reference values $A_0 = i_0 = \alpha_0 =1$ yields
\begin{eqnarray}
{\rm Im}\, \chain{ 111 i 1} &\equiv& {\rm Im}\, W_{i1} V^*_{i1} V_{11} W^*_{11} , \ 2 \le i \le \ng .
\label{newchain1}
\end{eqnarray}
These determine the $\ng-1$ phases in $\bar \Phi$.

We now summarize how the general high energy invariants can be reduced to those discussed above, using a by now familiar procedure. In loops and chains, any of the internal line indices of the same type cannot be repeated, because then the graph can be broken up into smaller invariant subgraphs by reconnecting the lines with the same label.  For loops, this decomposition rule is given by Eq.~(\ref{idone}).  The additional chain decomposition identities are
\begin{eqnarray}
\chain{B j \beta [k \ldots \gamma] k \alpha i A} &\to& \chain{B j \beta k \alpha i A } \loop{[k \ldots \gamma]}\nn
\chain{B j \beta [k \ldots \beta] l \alpha i A} &\to& \chain{B j \beta l \alpha i A } \loop{[k \ldots \beta]}
\label{chainid}
\end{eqnarray}
for repeated labels $k$ and $\beta$, respectively. 
A particularly useful identity obtained from Eq.~(\ref{chainid}) is
\begin{eqnarray}
\chain{B j \beta k \alpha i A } \loop{j \alpha}
&=& \chain{B j \alpha i A} \loop{ j \beta k \alpha}
\label{24b}
\end{eqnarray}
which replaces three matrix elements $V V^* V$ by a single matrix element $V$ in a chain. Thus, in large chains, one can apply Eq.~(\ref{24b}) to split off $VV^*VV^*$ bubbles.  Eq.~(\ref{24b}) is the generalization of Eq.~(\ref{24}) from loops to chains.

There also are reconnection identities.  For loops, these identities are Eq.~(\ref{idtwo}).  The chain reconnection identities are
\begin{eqnarray}
\chain{B j \alpha i A} \chain{D l \alpha k C } &=&
\chain{B j \alpha k C }\chain{D l \alpha i A}.
\end{eqnarray}

The identities imply that the invariants are $VV^*$ and $VV^*VV^*$ loops, and the $\eta_N$ invariant  $WW^*$ chains and products, and $WV^*VW^*$ chains,
\begin{eqnarray}
\loop{i \alpha}, &\quad&
\loop{i \alpha j \beta},\nn
\chain{A i A }, &\quad& \chain{AiB}\chain{BjA},\nn
\chain{AiB}\chain{AjB}, &\quad& \chain{A j \alpha i A }.
\end{eqnarray} 
The identities
\begin{eqnarray}
\chain{B j  \alpha i A } \chain{A_0 j A_0 } \chain{ A_0 i A_0 } &=& 
\chain{B j A_0}\chain{A_0 i A }\chain{A_0 j \alpha i A_0 } \nn
\chain{A_0 j \alpha i A_0 }\chain{A_0 j_0 \alpha j_0 A_0 } &=&
\chain{A_0 j \alpha j_0 A_0 }\chain{A_0 j_0 \alpha i A_0 }\nn
 &=& \chain{A_0 j \alpha j_0 A_0 }\chain{A_0 i \alpha j_0 A_0 }^* \nn
\chain{B j \alpha i A } \loop{j \beta i \gamma_0} &=& \chain{B j \gamma_0 i A} \loop{j \beta i  \alpha}
\end{eqnarray}
and Eqs.~(\ref{22}) and~(\ref{30}) imply that we need only consider the invariants
\begin{eqnarray}
\loop{i \alpha},\  \nn
\loop{i \alpha i_0 \alpha_0 }, \nn
\chain{A i A }, \nn
\chain{A_0 i A} \chain{A i_0 A_0}, \nn
\chain{A_0 i_0 A}^2,\nn
\chain{A_0 j_0 \gamma_0 i A_0}, 
\end{eqnarray}
and their complex conjugates, for fixed reference values $i_0,j_0,\alpha_0,\gamma_0,A_0$.  The first five of these invariants have already been studied in the sections on CKM and PMNS invariants, and the last one is the invariant in Eq.~(\ref{newchain}).

Choosing the reference values $\alpha_0=\gamma_0= A_0=i_0=j_0=1$ gives the independent invariants
\begin{eqnarray}
\loop{i \alpha},&&\qquad 1 \le i, \alpha \le \ng-1,\nn
{\rm Im}\, \loop{i \alpha  1 1 },&&\qquad 2 \le i, \alpha \le \ng-1, \nn
\chain{A i A },&&\qquad 1 \le i, A \le \ng-1,  \nn
{\rm Im}\, \chain{ 1 i A} \chain{A 11},&&\qquad 2 \le i, A\le \ng-1, \nn
{\rm Im}\, \chain{1 1 A}^2, &&\qquad 2 \le A \le \ng,\nn
{\rm Im}\, \chain{1  1 1 i 1},&&\qquad  2 \le i  \le \ng -1.
\end{eqnarray}

One can define a rephasing-invariant unit vector $v$, built from $\chain{A_0 j_0 \gamma_0 i A_0}$, 
\begin{eqnarray}
v_i &=& \frac{ \chain{A_0 j_0 \gamma_0 i A_0} }{\sqrt{ \loop{ j_0 \gamma_0}
\loop{i \gamma_0}\chain{A_0 j_0 A_0}}},
\end{eqnarray}
in addition to the matrices $T$ of Eq.~(\ref{tmatrix}) and 
\begin{eqnarray}
S_{i A} &=& { {\chain{A_0 i A} \chain{A i_0 A_0}} \over \sqrt{\chain{A_0 i_0 A_0}\chain{A_0 i A_0}\chain {A i_0 A} }}\ ,
\end{eqnarray}
the analogue of Eq.~(\ref{smatrix}).
The unit vector $v$ is equal to
\begin{eqnarray}
v_i &=& W_{i A_0} e^{i \phi(V_{j_0 \gamma_0})-i \phi(V_{i \gamma_0}) - i \phi(W_{j_0 A_0})}.
\end{eqnarray}
With $\mathcal{V}$ chosen so that the first row and column are real and non-negative,
and $A_0=j_0=\gamma_0=1$, the phase of $v_i$ is $-\bar \Phi_i$, so that $v$ fixes $\bar \Phi$.

\subsubsection{$\ng=2$}

$V$ has one angle, $W$ has one angle and one phase, and $\bar \Phi$ has one phase, for a total of four parameters, of which two are $CP$-even and two are $CP$-odd.
There are four independent invariants: $\loop{11}$, $\chain{111}$, ${\rm Im}\, \chain{112}^2$ and ${\rm Im}\, \chain{12111}$, which determine $\theta^{(V)}$, $\theta^{(W)}$, $\psi^{(W)}$ and $\bar \phi$, respectively.

\subsubsection{$\ng=3$}

The independent invariants involving only $V$ matrix elements are the $CP$-even 
\begin{eqnarray}
\loop{11} &=& |V_{11}|^2, \nn
\loop{12} &=& |V_{12}|^2, \nn
\loop{21} &=& |V_{21}|^2, \nn
\loop{22} &=& |V_{22}|^2, 
\end{eqnarray}
and the $CP$-odd
\begin{eqnarray}
{\rm Im}\, \loop{ 2211 } &=& {\rm Im}\,  V_{11} V^*_{12} V_{22} V_{21}^* ,
\end{eqnarray}
which determine all of the parameters $\theta^{(V)}_{1,2,3}$ and $\delta^{(V)}$ of $V$ in canonical CKM form.

The independent invariants involving only $W$ matrix elements are the $CP$-even
\begin{eqnarray}
\chain{ 111} &=& |W_{11}|^2, \nn
\chain{121} &=& |W_{21}|^2, \nn
\chain{212} &=& |W_{12}|^2, \nn
\chain{222} &=& |W_{22}|^2, 
\end{eqnarray}
and the $CP$-odd
\begin{eqnarray}
{\rm Im}\, \chain{122}\chain{211} &=& {\rm Im}\, W_{11} W_{12}^* W_{22} W_{21}^*, \nn
{\rm Im}\, \chain{112}^2 &=& {\rm Im}\, \left( W_{12} W_{11}^* \right)^2, \nn
{\rm Im}\, \chain{113}^2 &=& {\rm Im}\, \left( W_{13} W_{11}^* \right)^2,
\end{eqnarray}
which determine all of the parameters $\theta^{(W)}_{1,2,3}$, $\delta^{(W)}$ and $\psi^{(W)}_{2,3}$ of $W$ in canonical PMNS form.

There are two additional phases $\bar \phi_2$ and $\bar \phi_3$ which are determined by the $CP$-odd invariants
\begin{eqnarray}
{\rm Im}\, \chain{ 11121} &=& {\rm Im}\, W_{21} V_{21}^* V_{11} W^*_{11}, \nn
{\rm Im}\, \chain{ 11131} &=& {\rm Im}\, W_{31} V_{31}^* V_{11} W^*_{11}, 
\end{eqnarray}
respectively.

\section{Conclusions}\label{sec:conclusions}

We have determined all independent rephasing invariants of the quark and lepton mixing matrices in the seesaw model extension of the standard model and in its low-energy effective theory. In both theories, the independent rephasing invariants involving the quark CKM matrix $V$ are the quadratic invariants $\loop{i \alpha} = |V_{i \alpha}|^2$, $1 \le i, \alpha \le \ng-1$ and the imaginary parts of the quartic invariants $\loop{i \alpha 11 } = V_{i \alpha} V^*_{i_0 \alpha} V_{i_0 \alpha_0} V^*_{i \alpha_0}$, $2 \le i < \alpha \le \ng -1$.  The quadratic invariants determine the CKM matrix up to discrete ambiguities,  which are removed by the independent quartic invariants.

The lepton sector of the low-energy effective theory contains a single lepton mixing matrix, the PMNS mixing matrix $U$.  The independent invariants are $\uchain{ i \alpha i} = |U_{\alpha i}|^2$, $1 \le \alpha, i \le \ng-1$, and the imaginary parts of the quartic invariants $\uchain{1 \alpha i} \uchain{i 11} = U_{\alpha i} U^*_{\alpha 1} U_{11} U^*_{1 i}$,  $2 \le \alpha \le i \le \ng-1$, and $\uchain{11 i}^2 = \left(U_{1 i} U^*_{11}\right)^2$, $2 \le i \le \ng$. The discrete $\eta_\nu$ invariance  of the Majorana mass matrix requires that the phase in Eq.~(\ref{Ustd}) is $\Psi/2$ to maintain the $[0,2\pi)$ range for $\psi_i$.

The lepton sector of the high-energy theory contains two lepton mixing matrices $V$ and $W$.  The independent invariants involving only the lepton mixing matrix $V$ are the same set as for the quark CKM matrix.  The independent invariants involving only $W$ matrix elements are the same set as for the PMNS matrix.  The independent invariants involving both $V$ and $W$ matrix elements are the imaginary parts of  $\chain{111 i 1} = W_{i 1} V^*_{i 1} V_{11} W^*_{11}$, $2 \le i \le \ng$.

\medskip

AM would like to thank N. Wallach for discussions.

\end{document}